\documentclass[12pt,preprint]{aastex}
\begin{document}
\def\beq{\begin{equation}}
\def\eeq{\end{equation}}
\def\bey{\begin{eqnarray}}
\def\eey{\end{eqnarray}}
\def\pc{\, {\rm pc} }
\def\kpc{\, {\rm kpc} }
\def\msun{M_\odot}
\def\sun{\odot}
\def\lsim{\mathrel{\raise.3ex\hbox{$<$\kern-.75em\lower1ex\hbox{$\sim$}}}}
\def\gsim{\mathrel{\raise.3ex\hbox{$  $\kern-.75em\lower1ex\hbox{$\sim$}}}}
\def\Msun{M_\odot}
\def\Lsun{L_\odot}
\def\lsun{L_\odot}
\def\kms{\, {\rm km \, s}^{-1} }
\def\eV{\, {\rm eV} }
\def\keV{\, {\rm keV} }
\def\dis{{\rm dis}}
\def\grad{{\bf \nabla}}
\def\HSZ{{\it HSZ}}

\title{Coincidences of Dark Energy with Dark Matter -- Clues for a Simple Alternative?}
\author{HongSheng Zhao\altaffilmark{1}\\Scottish University Physics Alliance, University of St Andrews, KY16 9SS, UK}
\altaffiltext{1}{PPARC Advanced Fellow, hz4@st-andrews.ac.uk}
\begin{abstract}
A rare coincidence of scales in standard particle physics 
is needed to explain why $\Lambda$ or the negative pressure of cosmological dark energy (DE) coincides
with the positive pressure $P_0$ of random motion of dark matter (DM) in bright galaxies.
Recently Zlosnik et al. (2007) propose to modify the Einsteinian curvature by adding a non-linear pressure  
from a medium flowing with a four-velocity vector field $U^\mu$.
We propose to check whether a smooth extension of GR with 
a simple kinetic Lagrangian of $U^\mu$ can be constructed, and whether the pressure can bend space-time sufficiently to replace the roles
of DE, Cold DM and heavy neutrinos 
in explaining anomalous accelerations at all scales.
As a specific proof of concept 
we find a Vector-for-$\Lambda$ model (${\mathbf V\!\Lambda}$-model) and its variants.  With essentially {\it no free parameters},  these appear broadly consistent with the solar system, gravitational potentials in dwarf spiral galaxies and the bullet cluster of galaxies, early universe with inflation, structure formation and BBN, and late acceleration with a 1:3 ratio of DM:DE.
\end{abstract}

\maketitle

The incompleteness of 
standard physics and Einstein's General Relativity (GR) is evident from 
the smallness of the cosmological constant $\Lambda$ or the vacuum energy density 
${\Lambda c^2 \over 8 \pi G} \sim  (0.001{\rm eV})^4$, compared to 
the expected quantum pressure  
${c^5 m_P^4 \over \hbar^3} \sim (10^{28}{\rm eV})^4$ at scales of 
the Planck mass $m_P=\sqrt{\hbar c \over G}$.
Current speculations of the new physics of $\Lambda$
are as free as analogous speculations of the 
Pioneer Anomaly (Turyshev, Nieto, Anderson 2006), both represent acceleration discrepancies of order 
$\sim 7a_0$, driven by unidentified (likely unrelated) pressures $\sim 72P_0$, where $a_0 \equiv 1.2\AA/\sec^{2}$, and  $P_0 \equiv {a_0^2 \over 8 \pi G}$ are scales of acceleration and pressure.
On intermediate scales, galaxy clusters and spiral galaxies often 
reveal a discrepant acceleration of order $(0.1-2)a_0$.
GR, if sourced primarily by baryons and photons with negligible mass density of neutrinos and other particles in the Standard Model or  variations, 
appears an adequate and beautiful theory in the inner solar system, but
appears increasingly inadequate in accounting for astronomical observations 
as we move up in scales from 100AU to 1 kpc to 1Gpc.  
The universe made of known material of positive pressure
should show a de-accelerating expansion as an open universe, 
but instead it is turning into accelerating now, evidenced by  
much dimmer supervonae detected at redshift unity. 
A standard remedy to restore harmony with GR and fit successfully large scale observations 
(Spergel et al. 2006 and references therein) 
is to introduce a "dark sector", in which
two exotic components dominate the matter-energy
budget of the Universe at the redshift $z$ with a split of
$\Omega_{DE}:\Omega_{DM}=3:(1+z)^3$ approximately:
a Dark Energy (DE) as a negative pressure and nearly
homogeneous field described by unknown physics, and 
a Cold Dark Matter (DM) as a colissionless and pressureless
fluid motivated by perhaps the MSSM physics.   
However, anticipating several new particles from the LHC, 
the success of this Concordance Model still gives little clue to the physics of governing the present $1:3$ ratio of its constituents.   
This ratio is widely considered improbable, because standard particle physics expects a ratio $1:10^{120}$.  
Here we speculate whether the $3:(1+z)^3$ ratio could come from a coincidence
of scales of $a_0 \equiv 1.2\AA/\sec^{2}$ with 
the cosmological baryon energy density 
$\rho_b c^2 \sim 3.5 \times {(1+z)^3P_0}$.

{\it A deeper link of DM and DE}:
It is curious that the distribution of DM in dwarf galaxies is extremely ordered,
something that the cuspy $\Lambda$CDM halos  
are still struggling to explain even with the maximum baryonic feedback (Gnedin \& Zhao 2002).  
For example, on galaxy scales the Newtonian gravity of DM $g_{DM}={V_c^2 \over R}-g_B$ and Newtonian
gravity of baryons $g_B={GM_B \over R^2}$ have a tight correlation:
\beq\label{ggb}
(g^2/g_B)^{n} -g^n \approx a_0^n, \qquad g \equiv g_{DM}+g_B,
\eeq
where $n \ge 1$ (Zhao \& Famaey 2006).
This rule holds approximately at {\it all radii} R of 
all spiral galaxies of baryonic mass $M_B(R)$ and circular velocity $V_c(R)$
within the uncertainty of the stellar mass-to-light and object distance.
For low surface brightness galaxies or at the very outer edge of bright spirals,
the gravity $g$ is weaker than $a_0$, our empirical formula predicts 
$g^2/g_B = (V_c^2/R)^2/(GM_B/R^2) = V_c^4/(GM_B) \sim a_0$, 
which is essentially the normalisation of (baryonic) Tully-Fisher relation (McGaugh 2005).  
Bulges and central part of elliptical galaxies are dominated by baryons inside 
a transition radius where the baryon and DM contribute
about equally to the rotation curve, Eq.(\ref{ggb}) predicts $g_{DM}=g_B=a_0/2$; 
we can define a DM pressure $P_0 \equiv a_0 \times {a_0 \over 8\pi G}$
at transition by multiplying the local gravity $(g_{DM}+g_B)=a_0$ with  
the DM column density above this radius ${g_{DM} \over 4\pi G}=a_0/(8 \pi G)$.
This scale $P_0$ appears on larger scales too. 
All X-ray clusters have gas pressure and the DM random energy density comparable to $P_0$.
The amplitude of the scale $a_0$ appears in the $r^{-1}$ cusp of 
CDM halos too (Xu, Wu, Zhao 2007, Kaplinhat \& Turner 2001).
These can be understood since the last scattering shell at $z=1000$ has a thickness 
$2 L \sim 10$Mpc and contains typical potential wells of depth $c^2/N \sim (1000\kms)^2$ 
due to inflation, where $N \equiv 10^5$, hence the typical 
internal acceleration is $c^2/N/L \sim 0.2 a_0$.
Also a DM sphere of radius 5Mpc turning non-linear now would fall in with
an acceleration $\sim 200 \times H_0^2 \times 5Mpc\sim a_0$.
While correlations of baryon and DM can generally be understood in a galaxy formation
theory where DM and baryons interact, the unlimited freedoms of dark particles
means a good spread of its concentration, hence the correlation would have 
substantial history-dependent variance from galaxies to galaxies and radii to radii.  
For example, DM is unexpected in Tidal Dwarf Galaxies, is observed for its $a_0$
acceleration (Gentile et al. 2007).  
The tightness of such hidden regulations on DM at all radii for all 
galaxies is anomalous, at least challenging in the standard framework.  

It is even more curious that DM in various systems and DE
are tuned to {\it a common scale $P_0$, hence requiring a coincidence in two dark sectors.  
These empirical facts are unlikely random coincidences} of 
the fundamental parameters of the dark sectors.  
Since all these anomalies are based on the gravitational acceleration  
of ordinary matter in GR, one wonders if the dark sectors are not 
just a sign of an overlooked possible field in the gravitational sector.

Continue along Zhao (2006), 
here we propose to investigate whether the roles of both DM and DE 
could be replaced by a vector field in a modified metric theory.  
This follows from two long lines of investigations 
pursued by Kostelecky, Jacobson, Lim and others on consequences of 
symmetry-breaking in string theory, 
and by Milgrom, Bekenstein, Sanders, Skordis and others 
driven by astronomical needs.  These two independent lines were first merged by 
the pioneering work of Zlosnik et al.(2007).  {\it Existence of an explicit Lagrangian} 
satisfying main constraints for the solar system, galaxy rotation curves 
and cosmological concordance ratio remains to be demonstrated.

{\it Warming up to vector field}:
In Einstein's theory of gravity, 
the slightly bent metrics for a galaxy in 
an uniform expanding background set by the flat FRW cosmology is given by
\begin{equation}\label{metric}
g_{\mu\nu} dx^{\mu}dx^{\nu} 
=-(1+{2\Phi \over c^2}) d(ct)^2 + (1-{2\Psi \over c^2}) a(t)^2 dl^2 
\end{equation}
where $dl^2=\left(dx^2 +dy^2 + dz^2 \right)$
is the Euclidian distance in cartesian coordinates.
In the collapsed region of galaxies, the metric is quasi-static with 
the potential $\Phi(t,x,y,z)=\Psi(t,x,y,z)$ 
due to DM plus baryon, which all follow the geodesics of 
$g_{\mu\nu}$.   

Modified gravity theories are often inspired to preserve the Weak Equivalence Principle, 
i.e., particles or small objects still go on geodesics of above physical metric 
independent of their chemical composition.  Unlike in Einstein's theory, 
the Strong Equivalence Principle and CPT can be violated by, e.g., creating 
a preferred frame using a vector field.
The Einstein-Aether theory of Jacobson \& Mattingly (2001) 
is such a simple construction, where a unit vector field $U^{\mu}$  
is designed to couple only to the metric but not matter directly.
It has a kinetic Lagrangian with linear superposition of 
quadratic co-variant 
derivatives $\nabla (c^2U) \nabla (c^2U)$, where $c^2U^{\mu}$
is constrained to be a time-like four-momentum vector per unit mass by
$-g_{\mu\nu}U^{\mu} U^\nu= 1.$
The norm condition means the vector field introduces up to 3 new degrees of freedom; e.g., a perturbation in the FRW metric (Eq.\ref{metric}) has
$c^2U_{\mu} \equiv g_{\mu\nu}c^2 U^{\nu} \approx (c^2+\Phi,{A_x \over c},{A_y \over c},{A_z \over c})$, 
containing a four-vector made of an electric-like potential $\Phi$ and three new magnetic-like potentials.  But for spin-0 mode perturbations with a wavenumber vector ${\mathbf k}$, we can approximate 
$U_{\mu} - (1,{\mathbf 0}) \approx ({\Phi \over c^2},{{\mathbf k} V \over c})$, which contains 
just one degree of freedom, i.e., the flow potential $V(t,x,y,z)$.
We expect an initial fluctuation of $c|{\mathbf k}|V \sim |\Phi| \sim c^2 N^{-1} \equiv 10^{-5}c^2$ 
can be sourced by a standard inflaton; the vector field tracks the spectrum of metric perturbation (Lim 2004).  

Most recently Zlosnik et al. (2007) suggested to replace the linear $\lambda \nabla U \nabla U$ with 
a non-linear kinetic Lagrangian $ F(\lambda \nabla U\nabla U)$ to extend Jacobson's framework.
They showed this class of non-linear models is promising to produce 
the DE effect in cosmology and the DM-like effect in the weak field limit.  
Here we continue along the lines of the pioneering authors, but aim for  
a single Lagrangian with parameters in good match with basic observations of a range of scales.

{\it A Simple Lagrangian for $\Lambda$}:
The difficulty of writing down a specific Lagrangian is that 
there are infinite ways to form pressure-like terms quadratic to 
co-variant derivatives of the vector field.  Simplicity is the guide when choosing gravity since GR plus $\Lambda$CDM largely works.  
Let's start with forming two pressure terms for any four-momentum-like
field $A^{\mu}$ with a positive norm 
$m c^2 \equiv \sqrt{-g_{\alpha\beta}A^{\alpha}A^{\beta}}$ by 
\begin{equation}\label{Z}
8\pi G {\cal J}(A) \equiv 
{1 \over 3} \left({\nabla_\alpha A^\alpha \over m}\right)^2,~
8\pi G {\cal K}(A) \equiv {\nabla_\parallel A^\alpha \over m} {\nabla_\parallel A_\alpha \over m}
\end{equation}
where the RHSs are co-variant with dimension of acceleration squared,
and $\nabla_\parallel=A^\alpha \nabla_\alpha$ or $\nabla_\alpha$ stands for the co-variant derivative with space-time coordinates 
along the direction of the vector $A$ or the dummy index $\alpha$ respectively.  
From these we can generate two simpler 
pressure terms $K$ and $J$ of the unit vector field $U^\alpha$ by
\begin{equation}\label{KJgalaxy}
\begin{array}{cllcll}
J \equiv {\cal J}(U) &\sim&0,                 & K \equiv {\cal K}(U) &\sim&{|\nabla \Phi|^2 \over 8\pi G}  ~\mbox{\rm in galaxies}\\
 &\sim&{3 c^2H^2 \over 8\pi G},               &   &\sim&0  ~\mbox{\rm in flat universe}
\end{array}\label{KJcosmo}
\end{equation}
where the approximations hold for $U^\alpha$ with negligible 
spatial components and nearly flat metric (Eq.\ref{metric}).
Note the $J$ and $K$ are constructed so that we can control 
time-like Hubble expansion and space-like galaxy dynamics {\it separately}.\footnote{ 
A full study should include space-like terms $8\pi G K_{12} \equiv 2g^{\alpha\beta} 
(c^2\nabla_\alpha U^\gamma) (c^2\nabla_\beta U_\gamma) - {2 \over 3}(c^2\nabla_\alpha U^\alpha)^2$
and $8\pi G K_{13} \equiv  2g^{\alpha\beta} 
(c^2\nabla_\alpha U^\gamma) (c^2\nabla_\beta U_\gamma) - 2 (c^2\nabla_\alpha U^\beta)(c^2 \nabla_\beta U^\alpha)$
which change details of structure formation, PPN, and gravitational waves, 
which are beyond our goal here.}
The $K$-term, with a characteristic pressure scale ${a_0^2 \over 8\pi G} =P_0$ in galaxies, is the key for our model.  The $J$-term, meaning critical density, has a characteristic scale $N^2P_0 \sim 10^{10} P_0$:
at the epoch of recombination $z=1000$ when baryons, neutrinos, and photons contribute 
$\sim (8,3,5) \times 10^9 P_0$ respectively to the term $J={3c^2 H^2 \over 8 \pi G}$; 
so the epochs of equality and recombination nearly coincide.

Now we are ready to construct our total action $S=\int d^4x |-g|^{1\over 2} {\cal L}$ 
in physical coordinates, where the Langrangian density
\begin{equation}\label{action}
{\cal L} = {R \over 16 \pi G} + L_m + L_J + L_K + (U_\nu U^\nu \!\! +1) L^m, ~
\end{equation}
where $R$ is the Ricci scalar, $L_m$ is the ordinary matter Lagrangian.
For the vector field part, $L^m$ is the Lagrangian multiplier for the unit norm and
we propose the new Lagrangian 
\begin{equation}\label{LJ}\label{LK}
L_J = \int_{0}^{J} \!\!      \lambda_N\left(\sqrt{|J| \over P_0}\right) d J, ~
L_K = \int_{\infty}^{K} \!\!  \lambda_n\left(\sqrt{|K| \over P_0}\right) d K, 
\end{equation}
where the non-negative continuous functions 
$\lambda_i(x) = \left[0, \lambda(x)-\lambda(N)\right]_{\rm max}, 
~\lambda(x)=\left(1+{x \over i} \right)^{-n} $, 
where the subscript $i=$ either $n$ or $N$.  
Incidentally, $n=0$ gives GR.  The cutoffs (e.g., with $n=\pm 1$) 
guarantee a bounded Hamiltonian with
kinetic terms $L_K$ and $L_J$ always bounded between $\pm N^2P_0$ 
(e.g., in a lab near Earth $K \sim (10^{13}-10^{22}) P_0 > N^2P_0$, so $L_K=0$).  The condition at the tidal boundary $K=J=0$ is well-behaved too (cf. eq.44-48 of Famaey et al. 2007 on Cauchy problem).
Note $1-{dL_K \over dK}>\mu_{min} \equiv (1+N/n)^{-n} \sim 10^{-15}$ and $1 - {dL_J \over dJ} > \mu_B \equiv (1+N/N)^{-n} \sim 2^{-3}$.

Taking variations of the action with respect to the metric and the vector field, 
we can derive the modified Einstein's equation (EE) and the dynamical equation for the vector field.  The expressions are generally tedious (Halle 2007), 
but the results simplifies in the perturbation and matter-dominated regime that interest us. 
As anticipated in Lim (2004) the $ij$-cross-term of EE yields $\Psi-\Phi=0$ for all our models, which means
incidentally twice as much deflection for light rays as in Newtonian.
As anticipated in Dodelson \& Liguri (2007), 
the $ti$-term EE can be casted into that of an unstable harmonic oscillator 
equation with a negative string constant 
$\dot{\dot{V}} + b_1 H \dot{V} - (1-\mu_B) b_2 H^2 V  = S(\Phi, \dot{\Psi})$ if $(1-\mu_B)>0$, 
so we expect that $H V$ tracks $\Phi$.  The tt-term of the EE takes the form
\begin{equation}\label{tt}
8 \pi G \rho = 3\mu_B H^2 + 2 \grad \cdot [(1-\lambda_n) \grad \Phi]  - \Lambda_0 - Q(\dot{\Phi},\dot{V},V) 
\end{equation}
where we approximated $1-\lambda_N(x) \sim 2^{-n}=\mu_B$ as a constant in matter dominated regime where $J < N^2 P_0$ and
the $Q$-term is zero for static galaxies and uniform FRW flat cosmology. 
So the tt-equation of Einstein reduces to the simple form
\begin{eqnarray}\label{poisson}
4 \pi G \rho &=& \grad^2\Phi - \grad \cdot \left[\lambda_n\left({|\grad \Phi|\over a_0}\right)\grad \Phi\right],~\mbox{\rm in galaxies} \\
{8 \pi G \bar{\rho} \over 3\mu_B} &=& H^2 - {\Lambda_0 \over 3\mu_B}, ~\mbox{\rm in matter-dominated FRW}
\label{hubble}
\end{eqnarray} 
Here the pressure from the vector field creates new sources for the curvature.
The term ${\grad(\lambda_n(x)\grad \Phi) \over 4\pi G}$ in the Poisson equation acts as if adding DM for quasi-static galaxies.  A cosmological constant in the Hubble equation is created by 
\begin{equation}\label{cosmocst}
{\Lambda_0 c^2 \over 8\pi G} = -\!\!\int_{\infty}^{0} \!\!\! \lambda_n(x) d(P_0x^2) \approx {2(nP_0)^2 \over (n-1)(n-2)} 
\end{equation}
 
{\it For binary stars and the solar system},  
$4 \pi G \rho  -  \grad^2\Phi \approx 0$ is true 
because the gravity at distances 0.3AU to 30AU from a Sun-like star is much greater than the maximum vector field gradient strength $Na_0$, so ${dL_K \over dK}=0$; in fact, $|\grad \Phi| \approx {G M_\odot \over r^2} \sim (10^9-10^5) a_0$, and the typical anomalous acceleration is $N a_0 \mu_{min} \sim 10^{-10}a_0$, well-below the current detection limit of $10^{-4}a_0$  (Soreno \& Jezter 2006).  This might explain why most tests of non-GR effects around binary pulsars, black holes and in the solar system yield negative results;  Pluto at 40 AU and the Pioneer satellites  at 100 AU might show interesting effects. Extrapolating the analysis of Foster \& Jacobson (2006), we expect GR-like PPN parameters and gravitational wave speeds in the inner solar system.

{\it Near the edges of galaxies}, we recover 
the non-relativistic theory of Bekenstein \& Milgrom (1984) with a function  
\beq\label{mu}
\mu(x)  \equiv  1-\lambda_n(x)
   \sim \mu_{min} + x, ~\mbox{\rm if $x={|\grad \Phi| \over a_0} \ll 1$}.
\eeq
Note that  $\mu(x) \rightarrow x$ hence rotation curves are asymptotically flat except for a negligible correction $\mu_{min} \sim 10^{-15}$.  
In the intermediate regime $x=1$ our function with $1-\lambda_n(x) \sim (0.55-0.6)$ for $n=2-5$ respectively.  Eq.(\ref{ggb}) argues that galaxy rotation curves prefer a relatively sharper transition than $\mu(x)=x/(1+x)=0.5$ at $x=1$ (Famaey, Gentile, Bruneton, Zhao 2007)
where we can identify $g_B/(g_{DM}+g_B) = \mu(x)$.  So our model should fit observed rotation curves.

{\it For the Hubble expansion:} the vector field creates cosmological constant-like term  
${\Lambda_0 c^2\over 8\pi G} \approx 9 P_0 $ below the zero-point of the energy density 
in the solar system because 
the zero point of our Lagrangian (Eq.\ref{LK}) is chosen at $N^2 P_0 \le K <+\infty$.
During matter domination,  the contribution of matter $8 \pi G \rho$ and $\Lambda_0$ to the Hubble expansion $H^2$ (Eq.\ref{hubble}) is further scaled-up 
because the effective Gravitational Constant $G_{eff}=G/\mu_B = 2^n G \ge G$ with GR being the $n=0$ special case.
\footnote{While Carroll \& Lim (2004) found a scale-down of G because they were interested in stable spin-0 modes with $(1-\mu_B)<0$ for a restricted class ($c_4=0 \ne c_1$)
of Jacobson's models, Dodelson \& Liguri (2006) argue that 
an unstable growth of the vector field is helpful to structure growth in many gravity theories. }  
Coming back to the original issue of the $3:1$ ratio 
of matter density to our cosmological constant, Eq.(\ref{hubble}) predicts that 
${\Lambda_0 c^2 \over 8\pi G \mu_B}: {\bar{\rho}_b c^2 \over \mu_B} 
\sim {9 P_0 \over \mu_B}: {4 (1+z)^3 P_0 \over \mu_B}$, which is close to the desired $3:(1+z)^3$ ratio.  
Adding neutrinos makes the explanation slightly poorer. 
So the DE scale is traced back to a separate coincidence of scale, 
i.e., the present 
baryon energy density $\bar{\rho}_b c^2 \sim 4 P_0$, where $P_0$
contains a scale $a_0$ for the anomalous accelerations on galactic scale.
Our model predicts that {\it DE is due to a constant of vacuum}, preset by the modification parameter $n$ of the gravity; $n=0$ gives GR.

In our model, the effective DM (the dog) follows the baryons (the tail) throughout the universal $(1+z)^3$ expansion with a ratio set by $n$.  
To fit the $\Lambda$CDM-like expansion exactly, we note
the Hubble equation for a flat FRW cosmology with vector field
and standard mix of baryons, neutrinos and photons 
${\Omega_b h^2 \over 0.02} \approx {\Omega_{\nu}h^2 \over 0.002}{0.07{\rm eV} \over m_\nu} \approx {\Omega_{ph} h^2 \over 0.000025} \sim 1$ yields at the present epoch
\beq\label{Omega}
{\Omega_b + \Omega_\nu + \Omega_{\rm ph} \over \mu_B} 
= 1-{\Lambda_0 \over 3\mu_B H_0^2} =  \Omega_{m}^{\Lambda CDM}
\eeq
The 2nd equality fixes $\mu_B^{-1}=2^{n}=(8-8.4)$ 
if we adopt $a_0/c \approx H_0/6 \approx 12$km/s/Mpc and 
$\Omega_{m}^{\Lambda CDM}=(0.25-0.3)$.
The 1st equality would predict an uncertain but very small neutrino mass 
$m_{\nu} \sim \pm 0.3$eV.

The BBN also anchors any modification to GR.
In the radiation-dominated era $|J| ={3c^2H^2 \over 8 \pi G} \gg N^2 P_0$, 
the dynamics is driven by
\beq\label{raddom}
{8 \pi G \rho} \approx 3H^2 - \Lambda_0 - \Lambda_{N}, ~\mbox{\rm in radiation-dominated FRW},
\eeq
where ${\Lambda_{N}c^2 \over 8\pi G}  = - \int_{0}^{\infty} \!\!\! \lambda_N(x) d(P_0 N^2 x^2) 
= -N^2P_0/8$ for $n=3$ is a finite negative number, much smaller than 
the radiation pressure $\sim (z/1000)^3N^2P_0$.    
So the early universe is GR-like, especially the Hubble parameter 
at the BBN, insensitive to the precise value of $N^2 P_0$.

Note a more general version of our ${\mathbf V}$ector-for-$\Lambda$ model has a Lagrangian
\beq\label{LKLJ}
L_K+L_J = \lambda_K {\cal K}(U\lambda_K^{1 \over {\cal N}}) +\lambda_J{\cal J}(U\lambda_J^{1 \over {\cal N}}) -P_0 {\cal V}(\lambda_K,\lambda_J),
\eeq 
with four vector degrees in $U\lambda_J^{1 \over {\cal N}}$ and the scalar $\lambda_K/\lambda_J$; it is optional to replace $U\lambda_K^{1 \over {\cal N}}$ with $U\lambda_J^{1 \over {\cal N}}$ to reduce the total freedom to 4 as in Bekenstein's TeVeS.
Our simple model is equivalent to the special case of two non-dynamical scalar fields $\lambda_K$ and $\lambda_J$ with ${1 \over {\cal N}} \sim {1 \over N} \rightarrow 0$, hence ${\cal K}={\cal K}(U)=K$ and ${\cal J}={\cal J}(U)=J$ (Eq.\ref{Z}).  The potential is smooth with $P_0 {\cal V}(\lambda_K,\lambda_J)= \int_{\mu_{\rm min}}^1 
\left[{\cal H}(\mu_{\rm min} + \lambda_K -\lambda) - {N^2 \over n^2} {\cal H}(\lambda-\lambda_J-2^{-n})\right] P_n d\lambda $, where $P_n \equiv (\lambda^{-{1 \over n}}-1)^2 n^2 P_0$ and ${\cal H}(y)$ is the Heaviside function of $y$.
A vector field $A_{\mu} \approx (mc^2 + m\Phi, mc{\mathbf A})$ with a mass scale $m$ has
a quantum degeneracy pressure limit $\sim {c^5 \over \hbar^3}m^4$.  
It is intriguing that our model suggests the existence of a zero point vacuum energy 
${\Lambda_0 c^2 \over 8 \pi G} \sim P_0 {\cal V}(1,1) \sim 9P_0 \sim (0.001{\rm eV})^4$.
And the (positive) radiation pressure at the epoch of baryon-radiation equality coincides
with the cutoff energy density $P_0 {\cal V}(0,0) \sim -N^2 P_0 \sim -(0.3{\rm eV})^4$, and 
the vacuum-to-cutoff energy density ratio $\sim 9/N^2 \sim 10^{-9}$ 
coincides with the cosmic baryon-to-photon or baryon-to-neutrino number ratio $\eta \sim 3 \times 10^{-10}$ 
due to a tiny asymmetry with antibaryons.   Can theories like quantum gravity and inflation 
explain {\it these coincidences}?  Understanding these might give clues to how the four-vector potential 
of photons decouples from the baryon current vector, and decouples 
from our E\&M-like vector field $A^\mu$ in spontaneous symmetry breaking 
in the string theory (Kostelecky \& Samuel 1989, Carroll \& Shu 2006, Ferreira et al. 2007).

{\it Massive neutrinos are optional} for our model because  
the $L_J$ term creates a massive neutrino-like effect in cosmology
without affecting galaxy rotation curves.  There are a few ways to create
the impression of a fluid of 2eV neutrino in clusters of galaxies as well
(Angus et al. 2007, Sanders 2005, Zlosnik et al. 2007).  E.g.,  a general Lagrangian with
${\cal N} \sim n$ would have new dynamical freedoms $\mu \equiv 1-\lambda_K$ and 
$1-\lambda_J$, which satisfy second order differential equations in time in galaxies, reminiscent of fluid equations for DM.
Then the Bekenstein-Milgrom $\mu$-function would acquire
a history-dependent non-local relativistic correction of order 
${c \over {\cal N} a_0 \tau} \sim 1$ if the temporal variation (relaxation) time scale $\tau$ of the scalar field $\lambda_K$ 
is comparable to the Hubble time.  
This dynamical correction is hard to simulate, but is most important at the tidal boundary
of (merging) systems where a condensate of the dynamical freedoms $\lambda_K$ and $\lambda_J$ 
oscillate rapidly, could in principle act as an extra DM source to explain some outliers to the Bekenstein-Milgrom theory, 
e.g., the merging bullet cluster with its efficient lensing and high speed (Angus \& McGaugh 2007).
A dynamical field $\lambda_J$ is desirable as an inflaton to seed perturbations (Kanno \& Soda 2004).

In summary, 
we demonstrate as a proof of concept that 
{\it at least one alternative} Lagrangian for the gravity (Eq.\ref{action},\ref{LKLJ}) can be 
sketched down to resemble the GR plus $\Lambda$CDM 
but with somewhat {\it less-fining} 
in terms of fitting several types of observations
{\it from dwarf spiral galaxies to the cosmic acceleration}.  
The keys are a zero-point pressure scale $P_0$ at the edge of galaxies, 
and a universal convergence source term 
${1-\mu_B \over 8\pi G} (c^2\nabla_\alpha U^\alpha)^2$ below the cutoff pressure $N^2P_0$, which is near the epoch of equality and the last scattering.  
However, the CMB should be sensitive to the $\mu_B \equiv 2^{-n}$ modification parameter.
\footnote{In radiation-dominated era, 
the perturbed Poisson equation for radiation is approximately $ 16\pi G \delta \rho  \approx 2|{\mathbf k}|^2 \Phi$,
at very short-wavelength (hence large $x \sim |{\mathbf k}|\Phi/a_0 \gg 1$)  , where $\Phi=\Psi$. 
But after recombination, $Q \sim 2 q |{\mathbf k}|^2\Phi \propto 2(1-\mu_B)|{\mathbf k}|^2\Phi$, resembling a dissipationless DM term $4 \pi G\delta\rho_{DM}$ to make the matter perturbation grow as 
$4 \pi G \delta{\rho} \approx  \left(1-q \right) |{\mathbf k}|^2\Phi$.}      
It should be feasible to falsify the present model and variations 
by simultaneous fits to the supernovae distances and the CMB.

HSZ acknowledges helpful comments from Anaelle Halle, 
Benoit Famaey, Tom Zlosnik, Pedro Ferreira, Constantinous Skordis, David Mota, Eugene Lim, Meng Su and the anonymous referee, and the support from KITP during the gravitational lensing program 2006.

{}

\end{document}